# Single photon emitters in hexagonal boron nitride: A review of progress

A. Sajid[1,2,3*], Michael J. Ford[1,4] and Jeffrey R. Reimers[1,4]

[1]*University of Technology Sydney, School of Mathematical and Physical Sciences, Ultimo, New South Wales 2007, Australia.*
[2]*CAMD, Department of Physics, Technical University of Denmark, 2800 Kgs. Lyngby, Denmark.*
[3]*Department of Physics, GC University Faisalabad, Allama Iqbal Road, 38000 Faisalabad, Pakistan.*
[4]*International Centre for Quantum and Molecular Structures and Department of Physics, Shanghai University, Shanghai 200444, China.*

**Abstract:** This report summarizes progress made in understanding properties such as zero-phonon-line energies, emission and absorption polarizations, electron-phonon couplings, strain tuning and hyperfine coupling of single photon emitters in hexagonal boron nitride. The primary aims of this research are to discover the chemical nature of the emitting centres and to facilitate deployment in device applications. Critical analyses of the experimental literature and data interpretation, as well as theoretical approaches used to predict properties, are made. In particular, computational and theoretical limitations and challenges are discussed, with a range of suggestions made to overcome these limitations, striving to achieve realistic predictions concerning the nature of emitting centers. A symbiotic relationship is required in which calculations focus on properties that can easily be measured, whilst experiments deliver results in a form facilitating mass-produced calculations.

1. **Introduction**

The discovery[1] in 2016 of single-photon emission (SPE) from defects in hexagonal boron nitride (h-BN) has inspired significant research, receiving 286 citations by September 2019. Excitement stems from the possibility that such SPE could be harnessed by nanophotonics[2-5] industries to deliver applications. Indeed, different types of SPEs are already significant in many leading quantum technologies[2,6] including: quantum sensing[7-10], quantum nanophotonics[3,11-14], quantum information processing[15-16], quantum communication[17-18] and metrology[19]. Already well-studied solid-state SPEs include[4]: the negatively charged nitrogen-vacancy ($N_V^{-1}$) centre in diamond[20-23], silicon carbide[5] (SiC), and $ZnO$[24]. The exploitation of the colour centres in h-BN demands the ability to control and tune their structural arrangements and spectroscopic properties. The aim of this Review is to focus efforts towards the determination of their chemical structure, through examination of critical experimental information and, in particular, the shaping and sharpening of theoretical and modelling techniques for their interpretation and the types of experiments needed to facilitate it.

Driving this, the discovery of SPE from the 2D materials e.g. h-BN, WSe2 etc. has opened a new arena of research attributable to their unique optical and structural properties[1, 25-32]. Intense luminescence, facilitated by the close proximity of emitters to the surface, makes them promising for high quantum efficiency applications. Other advantages offered by SPEs in 2D materials include the ease with which they may be coupled to waveguides and their compatibility with other 2D materials that may be present in some complex device. Particular advantages of SPE in h-BN include: emission in both the visible[1, 33-38] and UV regions[39-41], brightness, controllability of emission polarization, and tunable emission[42], making them particularly interesting for quantum sensing and optical communications.

Defects either occur naturally in semi-conductor materials owing to unintentional incorporation of impurities or vacancies during crystal growth processes. In addition, they can be deterministically introduced through electron beam radiation[43-44] etc. In either case, defects have considerable effects on the electronic, magnetic and optical properties of host materials. Existence of fluorescent point defects at low enough density can enable a few SPEs to be isolated and be stable at room temperature. This happens when at least one of the electronic states responsible for emission is spatially localised, for example if it involves a defect level that is energetically far removed from bulk valence and conduction levels of the host lattice.

In bulk form, h-BN has an indirect band gap of 6.08 eV[45]. The band gap in monolayer or few-layer h-BN flakes has not yet been determined experimentally, but a recent theoretical calculation based upon an accurate version of Density Functional Theory (DFT) predicted the monolayer to have a direct band gap of 6.47 eV, with a cross-over to indirect band gaps occurring at two layers[46].

 Two different types of emission have been observed in h-BN. In one, the emission energy is in the 4-6 eV region in the UV, close to the h-BN band gap. Indeed, h-BN has shown promise as an ultra-violet emission source[14], with evidence of lasing at UV wavelengths[47]. Above band-gap excitation has revealed a number of photoluminescence features, with a strong luminescence peak in the UV observed at 5.76 eV[47] that is assigned to recombination of free excitons generated by the high-energy excitation. The spatial distribution of this luminescence is homogeneous both in crystallites[48-49] and in few-layer flakes[50-51], as measured by spatially resolved cathodoluminescence. Additional luminescence bands centred at 5.5 eV and 4 eV[39, 48-49, 52-53] have been identified as being strongly localized near dislocations and boundaries[48-51] and have been assigned to defect states of yet underdetermined type. Recently, anti-bunching (confirmation of SPE) has been demonstrated for h-BN defect emission at 4.1 eV[40].

In this review, we focus mostly on the other type of emission. This occurs in the visible part of the spectrum in the vicinity of 2 eV, but the energies of different emitters can be spread over a considerable, continuous, spectral range. Individual spectra can also vary considerably in shape, with two near-extremum examples highlighted in Fig. 1. Both examples consist of a relatively sharp line separated from a smaller broader feature, typically 150-200 meV lower in energy. Based upon the shapes and relative intensities of these features, they are assumed to be zero phonon lines (ZPL) and associated phonon side bands (PSB), respectively, originating from a localised emission source, most likely involving at least one defect level lying deep within the semiconductor band gap. In the following, we will also use this terminology to describe spectral features, detailing also ways in which additional physical effects could, in principle, also contribute to spectroscopy and function. In the majority of cases, this photoluminescence in the visible part of the spectrum has been observed following below bandgap excitation in either bulk, few-layer or monolayer h-BN.

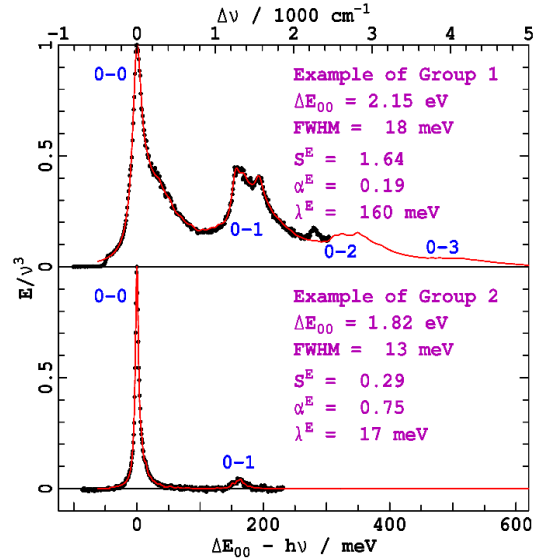

**Fig. 1** Extreme examples of Group-1 and Group-2 visible SPE observed from[33] h-BN defects (black points), and their fits to extrapolated Huang-Rhys models (red lines), using Lorentzian line profiles of specified FWHM. The fits are characterised by their total Huang-Rhys factors $S^E$, Debye-Waller factors $\alpha^E$, and reorganization energies $\lambda^E$, with spectral features labelled as the ZPL (0-0), the primary PSB (0-1), and overtone PSBs (0-2, 0-3, etc.). Data reproduced with permission from[33] for extracting of the various parameters.

Now many experiments have probed the characteristics of the visible SPE emission, including: strain tunability[42,54], high temperature stability[55], two photon non-linear excitation[56], resonant excitation[57], coupling to tapered optical fibre[58] and gold nano-spheres[59], spectrally tunable quantum efficiency[60] and photoluminescence (PL) up-conversion (anti-Stokes processes)[61]. They, in some cases, can possess a non-magnetic ground states[37], misalignment of the absorption and emission dipoles[34], spin-dependent inter-system crossing (ISC) between triplet and singlet manifolds[62], fast decaying intermediates and a long-lived metastable state accessible from the first excited electronic state[63]. Moreover, the deterministic creation and activation of some of the colour centres in h-BN has recently been demonstrated[64-65].

The best-understood SPE for defects in materials is the diamond nitrogen-vacancy centre in its negatively charged form, $N_V^{-1}$. It took over a decade to understand its underlying mechanisms, energy level structure, and excitation and decay pathways[20, 66]. Atomic-level calculations, generally within the framework of DFT, were a key ingredient to interpreting experimental results and progressing this understanding[21-22], along with thorough group-theoretical analyses[23, 67]. Despite this, there are still detailed features of the emission from this defect that are under debate, for example interaction with phonons[68]. The physics of SPE from defects in SiC is even less clear[5]. For SPEs in h-BN, a novel and significant feature is that defects are embedded in a 2D material, something with significantly different dielectric properties to those of traditional 3D materials[69].

Despite the now considerable literature, both experimental and computational on quantum emission from h-BN the atomic origin of this emission is still under debate. A recent experimental measurement combining electron paramagnetic resonance with photoluminescence spectroscopy provides strong evidence for the boron vacancy in its negative charge state as responsible for emission in the near-infra-red[70]. However a comprehensive understanding of the observed emission across a continuous range of almost 1 eV is still lacking. To enhance the contributions to this discussion from electronic structure calculations, methodologies are required that can calculate ground and excited-state energies to chemical accuracy, i.e., to better than 0.1 eV. This is extremely challenging for excited states (and ground states in some instances) as both dynamic and static electron correlation typically needs to be included in the calculations.

Although several colour centres have been proposed so far, despite the research efforts made to identify the origin of SPE in h-BN, further detailed investigations are needed to unveil the origin of this emission. Most experimental and theoretical efforts have aimed at understanding the photophysics of the emission, in order to exploit its potential in applications such as quantum technology and quantum sensing. We present an overview and a critical review of this body of literature in order to collect the varied work into a single reference source and aid researchers in unravelling the origin of this quantum emission. Important contradictions and similarities in the experimental literature about the properties of these SPEs are highlighted. Major computational/theoretical limitations and challenges are also discussed. We conclude with a range of suggestions and propose future efforts that are needed to reveal the structure of these SPEs, focusing also on how experimental investigations can be tailored to enhance the impact of modelling studies.

## 2. SPEs and their use in semiconductors and quantum technologies

### 2.1 Desired properties

The major challenge towards developing any quantum technology is the need to, in the one system, be able to externally address and control functionality, on the one hand, and to have the system evolve in a predictable way in isolation from its environment, on the other[71]. The role of SPEs in achieving such seemingly incompatible objectives has been reviewed elsewhere[4-5, 14, 72-73]. SPEs can be relevant in applications simply owing to the desirable properties of isolated photons. The key objectives can then be achieved through control of the interactions of the photons with matter and with themselves. Relevant requirements then include: the ability to couple the photons into devices, the ability to generate them reliably on demand, and the intrinsic photon properties themselves such as their purity, lifetime, and indistinguishability (absolute minimal energy distribution and polarization variability), difficult requirements to achieve. Alternatively, the photons produced by SPEs may be relevant to devices simply through their role in the initialization or readout of some materials-based quantum system, an example being optically detected magnetic resonance (ODMR)[74]. In ODMR, photo-excitation and photoluminescence is used to initialise and readout electronic spin states for a variety of applications including the ability to detect small and local variations in the ambient magnetic environment. Recent experimental results show promise of this capability in h-BN[70, 75].

Many basic properties of SPEs are desired, independent of the role of the photon in a device, as sketched in Fig. 2. Finding all desired features in one system is a major challenge, and currently no solid-state system possess all of these properties. Desired are: the need to operate at room temperature, week environmental interactions as these could cause dephasing and spectral diffusion, operation on demand, photostablility without blinking, and to emit photons with high efficiency and tightly controlled properties such as photon energy, polarization, lifetime, and the excited-state lifetime (the time required for their production following stimulation). An important associated quantitative measures is the quantum efficiency ϕ representing the fraction of emitted photons relative to the number of photons adsorbed. Values of ϕ ~ 1 indicate that the SPEs efficiently responds to stimulation, rather than dissipating the absorbed energy by other means. As Fig. 1 exemplifies, emitted light will always span some energy range, with typically the most intense emission being through the ZPL, with other phonons being emitted at lower energies through its associated PSB. The Deybe-Waller factor of the emitter, $\alpha$, is the fraction of photons emitted in the ZPL and therefore provides a key indication as to how reproducible the emission is and hence how useful the SPE would be for device applications[4-5]. The full-width at half maximum (FWHM) of the ZPL is a related key indicator.

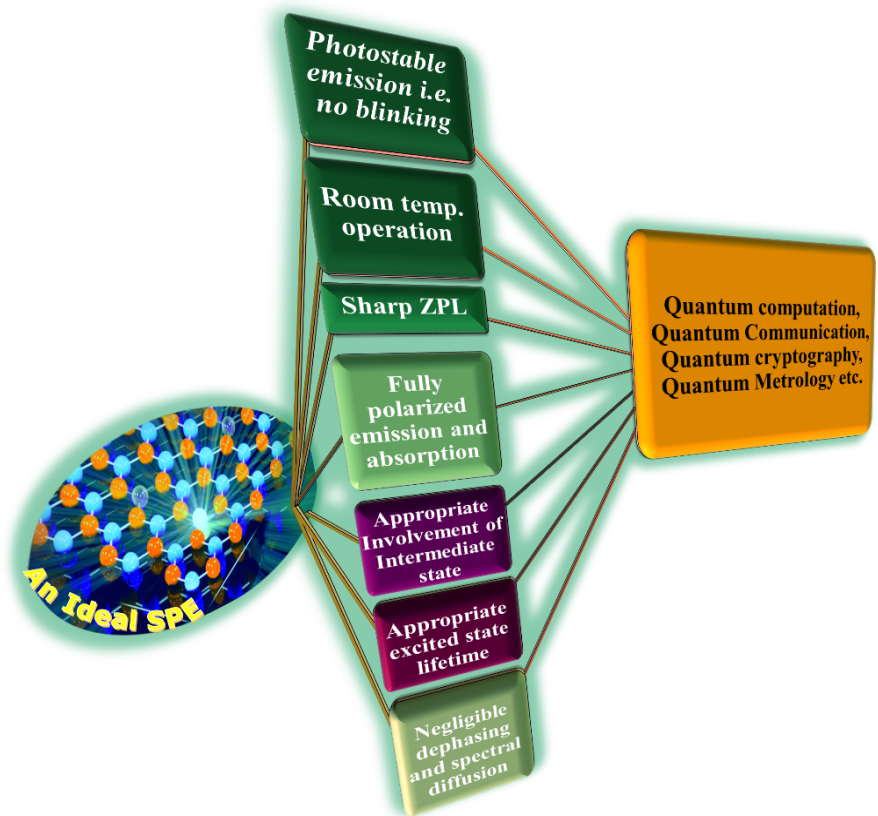

**Fig. 2** Important properties relevant to SPEs that could lead to applications in quantum technologies[4-5, 72]

## 2.2 Comparison of h-BN SPEs to other SPE types

A number of SPEs, including atoms[72], quantum dots[76] and superconducting circuits[77], have already been utilized in technologies. SPEs based on atoms or trapped ions and quantum dots have excellent optical properties[78-79]. Unfortunately, scalability and integration issues, in the case of atoms or trapped ions, or problems involving low coherence times and indistinguishability, in the case of quantum dots, has so far limited their usage in practical applications. Some of these issues are currently being investigated in detail[80].

SPEs in semiconductors offer alternative technologies, providing ease of integration with nanophotonic devices to facilitate quantum measurement. The application of point defect based SPEs for quantum technologies have already been demonstrated (to varying extents) in quantum computation[15-16], quantum communication[17-18] and metrology[19].

So far, the $N_V^{-1}$ centre in diamond is the most thoroughly studied semiconductor SPE whose crystallographic and electronic structure has largely been resolved. SPEs in diamond are photostable, operate at room temperature and have long coherence times. However the $N_V^{-1}$ centre has a Debye-Waller factor of only $\alpha = 3\%$[81], even at low temperature, which limits it's applicability for quantum technologies. The quantum efficiency of $N_V^{-1}$ centre in diamond is not optimal at 70%, owing to the existence of a metastable state that provides an alternative decay channel[81]. In addition, $N_V^{-1}$ has a non-zero local electric dipole moment in both its ground and excited states, making it sensitive to local electric fields including those arising from local strain, effects that contributes to undesirable homogeneous and non-homogeneous spectral broadening[4].

Another source of SPE, from the negatively charged silicon vacancy centre in diamond ($SiV^-$) that can be obtained by substituting two adjacent carbon atoms with a single silicon atom located at their midpoint, has local $D_{3d}$ point-group symmetry that embodies inversion symmetry. It has a high Debye-Waller factor of 70%[82], but it has a very low quantum efficiency of just 3.5%. Similarly defects in SiC have lower Debye-Waller factors of 33%, but much higher quantum efficiencies of 70%[83]. Extensive experimental and theoretical studies are being performed to explore new solid-state systems that are potential hosts of SPEs with optimal parameters.

Current evidence suggests that h-BN SPEs could be competitive in applications compared to these materials. A significant issue, however, is observed variability in h-BN SPE performance. Defects can have Debye-Waller factors as low as 15%[84], but, more encouragingly, in some circumstances up to 82% have been proposed[1]. Also, these defects typically show spectrally varying quantum efficiency in the range of 50-100%[60].

## 2. Observed properties of h-BN SPEs and the conceptual framework used to interpret them

### 2.1 Defect and lattice band structures.

An electrically driven point defect in a semiconductor with localized levels having energies lying inside the bandgap could be similar to the isolated atomic (quantum) systems developed historically for quantum applications. This situation is illustrated in Fig. 3 for h-BN. Orbital energy levels far removed from the conduction band (CB) and valence band (VB) of the semiconductor are called "deep levels" and cannot be accessed using just thermal energy. As deep levels can be highly isolated from their environment, SPE system based on semiconductors can offer attractive features such as long coherence times of the defect states formed after photo-illumination, allowing for extended quantum processing possibilities, room temperature operation, and reliable detection of qubit properties utilizing stable photoemission.

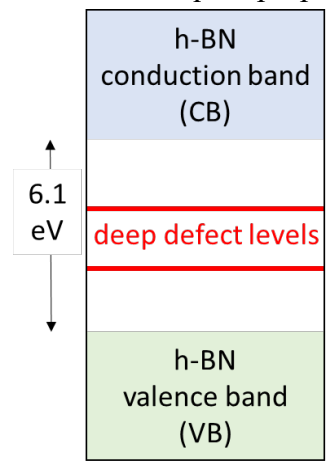

**Fig. 3** Energy-level diagram categorizing the one-electron energy levels of electrons in terms of those in the host valence and conduction bands, plus those in levels deep inside the host bandgap associated with material defects, showing the band gap of h-BN.

### 2.2 Proposed models for defect chemical structure and their nomenclature

Many different types of defects have been proposed[1, 37, 85-87,70, 75, 88-89,90]. To describe their chemical nature, a nomenclature system has been introduced that is appropriate to many materials[91]. In h-BN, defects usually involve vacancies in which one boron or nitrogen atom is absent; these are labelled "$V_B$" and "$V_N$", respectively. At such centres, net charge can be attracted, with labels then applied of, e.g., $V_N^{-1}$ (negative charge) and $V_N^{+1}$ (positive charge). Alternatively, an impurity atom could substitute for a lattice atom. If, e.g., a boron atom is replaced by nitrogen atom, the defect is labelled "$N_B$", but the replacement atom could be anything from the periodic table, e.g., "$C_B$" for a carbon impurity. It is also conceived that defects may not occur in isolation but instead be paired together on adjacent atomic sites. Hence the label "$V_N N_B$" indicates that there is a vacancy at a nitrogen site, with a boron neighbour replaced with nitrogen. Adjacent double vacancies have also been considered and are labelled "$V_{BN}$"[37].

All atoms neighbouring a vacancy will have broken chemical bonds in both their $\sigma$ and $\pi$ orbital systems. As three atoms surround each vacancy, this means that 6 defect-localized orbitals will be intrinsically involved. These orbitals interact with the surrounding h-BN CB and VB and will delocalize accordingly, with the extent of the delocalization controlled mostly by energy differences. Some of the 6 orbitals will themselves fall into the CB or VB, but others will be left to form the deep defect levels sketched in Fig. 3. The number of electrons that needs to be placed in these 6 orbitals in order to determine the critical electronic states of the defect depends on defect composition and charge. By this process, changing the chemical nature of the defect can have profound effects on spectral properties.

**2.3 Characterisation of observed emission in terms of electronic-state properties.**

Understanding of physics of defect emission requires an understanding of the electronic states involved and how they couple to atoms within the defect and the surrounding lattice. Various defect states arise through varying the occupancy of the deep defect levels sketched in Fig. 3, possibly also in combination with occupancy changes in either the CB or VB of the host semiconductor. These states can be represented by potential-energy surfaces, as sketched in Fig. 4. Absorption and emission of light transfers the defect from one state to another. Figure 4 shows, in particular, the effect that changes in the defect geometry have on the energy of the initial and final electronic states associated with photoluminescence. States can have different electron spin, depending on the number of spin-up and spin-down electrons they embody and the coupling between them. If the states have the same spin (e.g, both singlet or doublet states), then emission usually occurs rapidly (e.g., on the ps-ns timescale) and is known as *fluorescence*, whereas if they have different spins (e.g., triplet and singlet), then the emission will occur slowly (e.g., on the μs-s timescale) and is known as *phosphorescence*.

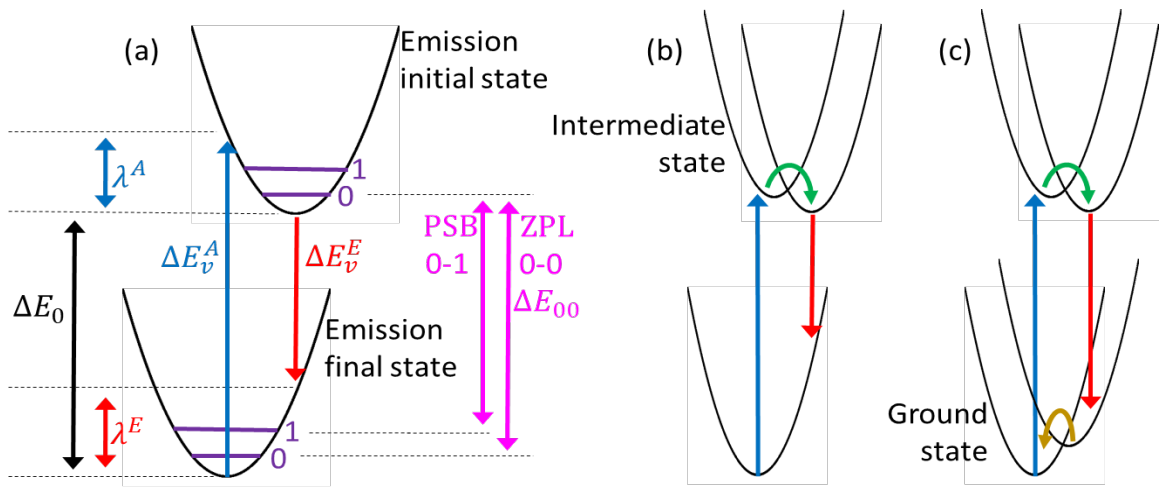

**Fig. 4** Sketches of potential-energy surfaces depicting different possible mechanisms for absorption followed by SPE. (a) Direct absorption (blue) followed by emission (red), with the initial and final electronic states for emission marked, as well as key vibrational levels and energies, including: the zero-point vibrational level with no phonons and a level with one phonon marked on each surface; the energy of the ZPL, $\Delta E_{00}$, and the reorganization energies released after vertical absorption and emission, $\lambda^A$ and $\lambda^E$, respectively. (b) Indirect absorption (blue) to an intermediate state, followed by either internal conversion or intersystem crossing (green) to the initial state for emission (red). (c) Indirect absorption (blue) followed, e.g., by intersystem crossing (green), emission (red) to the ground state of the second spin multiplicity, then recovery to the absolute ground state (brown).

The two states of the defect involved in the emission may or may not both be involved in the light absorption process. A wide range of possible scenarios can therefore arise, with three possibilities sketched in Fig. 4(a), Fig. 4(b), and Fig. 4(c). In Fig 4(a), only two states are involved in both the light absorption and light emission processes, known as *direct* absorption or resonant excitation. The other two scenarios involve *indirect* absorption to an intermediate state that then transfers its energy to the initial state for the emission. This energy transfer may proceed either adiabatically over a transition state or else non-adiabatically through violation of the Born-Oppenheimer approximation near some conical intersection. If the two states involved have the same spin multiplicity, then this energy transfer is called *internal conversion*, otherwise it is called *intersystem crossing*. The difference between the scenarios in Fig. 4(b) and Fig. 4(c) is that in 4(c) a fourth state is the final state of the emission, which then must subsequently recover back to the ground state in order to complete a full photocycle. Scenario 4(c) would most likely occur following intersystem crossing or else following some excited-state isomerisation of the defect, but intersystem crossing would seem the most relevant process for h-BN defects.

We provide the mathematical description most usually applied to interpret spectra associated with direct absorption, see Fig. 4(a). Incoming photons with average energy $\Delta E_v^A$ produce the excited state. Excess energy equal to the absorption reorganization energy $\lambda^A$ is then dissipated as heat through energy transfer into the vibrational modes of the h-BN lattice followed by the emission process of interest. This emission occurs at average energy $\Delta E_v^E$, followed by heat loss equal to the emission reorganization energy $\lambda^E$ and return to the original equilibrium conditions. The sum of the two reorganization energies $\lambda^A + \lambda^E$ tells the energy difference between the average absorption and emission energies $\Delta E_v^A - \Delta E_v^E$ and is known as the *Stokes shift*, with the average transition energies $\Delta E_v^A$ and $\Delta E_v^E$ usually referred to as *vertical* transition energies. They are readily calculable using DFT by evaluating the energies of both states at the two equilibrium geometries. The difference in energy between the two equilibrium geometries is called the *adiabatic* transition energy $\Delta E_0$. When corrected for the zero-point motion of each state, this gives the energy of the ZPL, $\Delta E_{00}$. This zero phonon line along with the phonon side bands associated with emission or adsorption are the quantities measured experimentally in photolumiscence or resonance excitation spectra. From experimental data, the reorganization energies are usually expressed as

$$\lambda^A = h \frac{\int_0^\infty A(v)dv}{\int_0^\infty A(v)/vdv} - \Delta E_{00}, \quad \lambda^E = \Delta E_{00} - h \frac{\int_0^\infty E(v)/v^2 dv}{\int_0^\infty E(v)/v^3 dv},$$

to provide the simplest measures of the spectral bandwidths.

Observed spectra are usually interpreted using the Huang-Rhys model[92]. This makes five assumptions concerning the properties of the electronic states involved. The first three of these are: (i) that the Born-Oppenheimer approximation[93-95] holds, (ii), that the Franck-Condon approximation holds[96] (i.e., no Herzberg-Teller effects[97]), and (iii), the potential energy surfaces can be described as harmonic functions expanded about the equilibrium geometries of the initial and final states. Based on these assumptions, spectra at 0 K can be characterized by their total Huang-Rhys factors

$$S^A = \frac{\int_{1-0 \text{ lines}} A(v-\Delta E_{00}/h)/vdv}{\int_{0-0 \text{ line}} A(v-\Delta E_{00}/h)/vdv}, \quad S^E = \frac{\int_{0-1 \text{ lines}} E(\Delta E_{00}/h-v)/v^3 dv}{\int_{0-0 \text{ line}} E(\Delta E_{00}/h-v)/v^3 dv}.$$

These factors tell the total number of phonons excited by vertical absorption and emission, respectively, and stem from partitioning the total absorption and emission into contributions arising from the ZPL (0-0) and parts of the PSBs associated with an increase of one phonon (0-1 in emission, 1-0 in absorption) of vibrational excitation (see Fig. 4(a)). Note that absorption and emission can also involve the excitation of overtone modes involving multiple phonons (0-*n* in emission, *n*-0 in absorption). Figure 1 includes a qualitative description of observed emission spectra into substructures identified based on line prominence. Note, however, that the 0-1 PSB extends all the way to zero frequency change, and therefore that overtones of low-frequency modes appear underneath and alongside the transitions associated with high-frequency 0-1 fundamentals. The total Huang-Rhys factors are simply related to the Debye-Waller factors $\alpha$ that tell the fraction of absorption/emission that occurs through the ZPL:

$$\alpha^A = \frac{\int_{0-0\,\text{line}} A(\nu - \Delta E_{00}/\hbar)/\nu \, d\nu}{\int_{-\infty}^{\infty} A(\nu - \Delta E_{00}/\hbar)/\nu \, d\nu} = e^{-S^A}, \quad \alpha^E = \frac{\int_{0-0\,\text{line}} E(\Delta E_{00}/\hbar - \nu)/\nu^3 \, d\nu}{\int_{-\infty}^{\infty} E(\Delta E_{00}/\hbar - \nu)/\nu^3 \, d\nu} = e^{-S^E}.$$

Unfortunately, determination of the reorganization energies and the Debye-Waller factors by spectral integration can be difficult as the above integrals extend over the entire band, which may be difficult to observe, and spectral features associated with the ZPL and PSB often overlap and cannot easily be separated out. Important features of the above equations is that they feature as critical elements the band-shape functions $\frac{A(\nu)}{\nu}$ and $\frac{E(\nu)}{\nu^3}$ depicting the material properties generating absorption and emission, respectively[98]. If, as is common practice, the emission is measured scanning the spectrometer linear in wavelength as $E(\lambda)$, then $E(\nu) = E(\lambda)/\nu^2$ and the emission band-shape is therefore $\frac{E(\lambda)}{\nu^5}$. In the SPE field, it is usual to present $E(\nu)$ in figures, even if $E(\lambda)$ is actually measured, so care must always be taken when considering published spectra. These scaling factors are important as they inhibit the reliable experimental determination of spectral tails, distorting bandshapes to allow them to become lost amongst experimental noise. As a result, the preferred method for determining critical Debye-Waller factors is often to extract first the Huang-Rhys factors $S^E$ from the spectra, a task that requires only processing of the 0-0 and 0-1 spectra regions. For this, the difficulty is that the 0-0 and 0-1 sub-bands can overlap, with 0-2, 0-3, etc. sub-bands also possibly interfering, so the task becomes one of separating out these varying spectral contributions.

In Fig. 1, two example SPEs are shown and analysed. Broadly, visible emitters in h-BN can be categorized into type types[33, 37], known as "Group 1" and Group 2", with few intermediary cases known. Group 2 emitters have strong ZPL 0-0 and weak 0-1 PSB contributions with little other emission, making determination of the Debye-Waller factors straightforward by both methods. However, the Group 1 emitters produce final states excited by many phonons, demanding sophisticated analyses of the ZPL and PSB.

In the Huang-Rhys model, spectra at 0 K are described in terms of a lineshape function for the ZPL and Huang-Rhys factors $S_i^E$ and $S_i^A$. The Huang-Rhys factors arise from the change in geometry of the material that occurs as a result of the electronic transition in terms of squared displacements projected onto each vibrational mode $i$ of the final electronic state, either $\delta_i^E$ for emission $\delta_i^A$ for absorption, where $S_i^E = (\delta_i^E)^2/2$ and $S_i^A = (\delta_i^A)^2/2$. Connection between the two sets of displacements, and hence the Huang-Rhys factors, is provided by[99] the Duschinsky rotation matrix[100] **D** which specifies the relationship between the forms of the normal modes of vibration in the two states. The total Huang-Rhys factors mentioned before are just these factors summed over all vibrational modes:

$$S^E = \sum_i S_i^E \text{ and } S^A = \sum_i S_i^A,$$

whilst the reorganization energies are just these weighted by the mode vibration frequencies:

$$\lambda^E = \sum_i h\nu_i^{GS} S_i^E \text{ and } \lambda^A = \sum_i h\nu_i^{ES} S_i^A.$$

where $\nu_i^{GS}$ and $\nu_i^{ES}$ are the vibration frequencies in the ground state and the excited state, respectively.

In defect SPEs, geometry changes upon electronic transition involve mostly the atomic bond lengths between atoms at the defect site. These give rise to the intense PSBs observed in the 1200-1600 cm$^{-1}$ (150-200 meV) spectral region, see Fig. 1, but they also interact with h-BN lattice vibrations and hence become distributed over the continuous vibrational degrees of freedom of the host lattice. The sums over discrete vibrational modes depicted above should therefore be replaced with integrals, but in practice such integrals are always evaluated at discrete points and so will use the summation notation throughout. If $m$ discrete vibrational modes are used, then each mode $i$ may be occupied by $n_i$ phonons represented by vector **n**, producing vibrational levels containing $\sum_i n_i$ phonons total. At this point, the remaining two assumptions used in the Huang-Rhys model need to be introduced: (iv) that the vibration frequencies are the same in the initial and final state, $\nu_i^{ES} = \nu_i^{GS}$ for all $i$, and that the Duschinsky matrix is simply **D** = **1**. The contribution to the bands shape from the 0-**n** emission transition is then given by the product

$$I_\mathbf{n} = \prod_{i=1}^m I(S_i^E, n_i)$$

where

$$I(S, n) = \frac{e^{-S} S^n}{n!}$$

are the squares of the Franck-Condon integrals[96]. General expressions for the spectral intensity based *only* on the Born-Oppenheimer approximation that evaluate all individual line intensities are known[101-102] and available in modern software packages[103]; with non-Born-Oppenheimer approaches also widely known[104] and commonly applied to modern spectroscopic challenges pertaining, e.g., to solar energy harvesting and transport[105-106]. Nevertheless, within the Huang-Rhys model, much simpler approaches become available in which the band shape is expressed more succinctly as the repeated convolution of the ZPL line-shape function $I_{zpl}$ with the intensity functions for each individual mode as

$$I = I_{zpl} \otimes I_1 \otimes I_2 \ldots \otimes I_n,$$

where $I_i$ is the spectrum made by combining all lines $I(S, n)$. Spectra determined from the Huang-Rhys model are invariant to density in that if one mode at frequency $\nu_i$ with Huang-Rhys factor $S_i$ is replaced by two modes, both at frequency $\nu_i$ with Huang-Rhys factor $S_i/2$, then the resulting spectrum is unchanged. This situation arises as, even though the individual integrals $I(S, n)$ scales exponentially with $S$, the total number of contributing vibrational levels **n** also scales exponentially and so the effects cancel. Discrete approaches that evaluate all individual line intensities are hence intrinsically numerically unstable, whereas the convolution approach is stable and linearly scales with the number of included vibrational modes. This can be done[99] very efficiently making only the Born-Oppenheimer approximation and a stable and reliable approximation to the Duschinsky matrix, including both Herzberg-Teller effects and treatment of large-amplitude anharmonic bending and torsional motions often predicted for proposed defects, and applied to data determined using many software packages including VASP[107] and Gaussian-16[103].

In Fig. 1, two observed spectra are interpolated using the convolution method, as implemented in the DUSHIN programme suite[99]. The result is the determination of *m* values of $S_i^E$ for each SPE, along with an extracted ZPL profile $I_{zpl}$, therein taken to be a Lorentzian function of determined full-width and half-maximum (FWHM). In the figure, the determined Huang-Rhys profiles are then extrapolated towards the band tail, highlighting parts of the emission spectrum that are difficult to observe experimentally but are nevertheless important to proper characterisation. Basic characterisation of Group 1 and Group 2 SPEs is summarized in Table 1. The challenges to providing such quantitative interpretations of experimental data pertain to: the quality of the treatment of the ZPL lineshape, assumptions made as to baseline, the spectral extent of the observed data, the possible presence of sharp emission from Group 1 defects appearing within the bands of Group 2 defects, especially within the main PSB, possible interactions between nearby defects, and failure of basic assumptions such as the Franck-Condon approximation, the Born-Oppenheimer approximation, or the form of the Duschinsky matrix. Directly predicting such quantitative data is one of the most important ways through which DFT and other electronic-structure calculations can contribute to determining the chemical nature of h-BN defect sites.

Table 1 Categorisation of visible SPEs in h-BN into two broad groups[33, 37, 85].[a,b]

| type | $\Delta E_{00}$ / eV | FWHM / meV | $S^E$ | $\alpha^E$ | $\lambda^E$ / meV |
|---|---|---|---|---|---|
| Group 1 | 1.8 – 2.2 | 0.08 – 0.35 | 0.9 – 1.9 | 0.15 – 0.4 | 60 – 160 |
| Group 2 | 1.4 – 1.8 | 0.04 – 0.08 | 0.25 – 0.6 | 0.5 – 0.8 | 15 - 30 |

a: Group 2 emitters have also been observed[37] with PSB intensities under 2% of the ZPL, suggesting that $S^E$ < 0.25 and $\alpha^E$ > 0.8.

b: see also refs.[1, 34, 38, 42, 55, 108-110].

## 2.4 The ZPL energy $\Delta E_{00}$ and temperature-dependent lineshape

Several groups across the globe have reported SPE from h-BN, with some common features in terms of energy profile. In this section, we review those common features and develop correlations. The first report on room temperature SPE from h-BN monolayers observed a ZPL energy of 1.99 eV, with the emission tentatively and contentiously assigned to the $V_NN_B$ defect[1].

Shortly after this, Group 1 and Group 2 SPEs (Table 1) were first categorized[33]. In that report, Group 1 emitters had broad and asymmetric ZPLs with energies in the range 1.8 to 2.2 eV, whereas Group 2 emitters had narrow and more symmetric ZPLs with energies in the range 1.6 to 1.8 eV[33]. Two Group 1 type emitters with ZPL energies at 1.82 eV and 2.16 eV were later studied for their spectral shift and temperature dependent line-width[111]. It was found that a 532 nm laser directly excited the SPE observed at 2.16 eV (Fig. 4(a)), while the 1.82 eV SPE was excited indirectly[111]. Initially it as suggested that the indirect mechanism involved cross-relaxation driven by excitonic coupling between defects, but no evidence supporting this was presented[111], with later works[34] instead suggesting that relaxation occurs on the same defect (Fig. 4(b)). The natural line-widths extracted from these measurements were found to be 0.352 and 0.148 μeV for the 2.16 eV ZPL and the 1.82 eV ZPL, respectively, and were independent of temperature. The amplitude of the 1.82 eV SPE was found to decrease much more rapidly with temperature compared to the 2.16 eV one[111]. Similarly, two types of emitters were reported in another work[37], one with a broad ZPL centred at 2.15 eV and the other with a sharp ZPL at 1.45 eV. The emitter with sharp ZPL was found to have almost no PSB[37]. Another group of researchers reported highly efficient and ultra-bright quantum emitters, of two types or families, in h-BN with ZPL centred at 1.97 eV and 2.08 eV[35]. Group 2 emitters have been identified near structural defects in the basal plane of the h-BN[37].

To eliminate possible substrate effects on emission from h-BN, SPEs in suspended single crystalline h-BN films have been formed with ZPL energies in the range 1.77 - 2.25 eV[38]. Single photon emission have also been demonstrated from h-BN powder and an exfoliated h-BN flake with ZPL in the energy range of 1.8 - 2.5 eV[112]. Room temperature SPE from a zero-dimensional boron nitride allotrope (the boron nitride nano-cocoon) has also been demonstrated with ZPL energy centred around 2.14 eV[110].

The lineshape of the ZPL is usually Lorentzian in basic shape, but this can undergo significant distortions and variation, especially for Group 1 SPEs. Complex lineshapes can be indicative of: more than one process controlling the excited-state lifetime, inhomogeneous broadening associated with ultra-fast spectral diffusion present in these emitters[113], hot-band anti-Stokes 1-0 transitions enabled by thermal phonon population in the initial state before emission[61], or interference from other emitters at slightly different frequencies. Whereas anti-Stokes transitions usually just broaden the low wavelength side of the ZPL in accordance with the appropriate Boltzmann factor, emission at energies of 160 meV higher than excitation can also be observed by this process, with the expected extremely low quantum yield[61].

Such effects may become important technologically as quantum emitters in h-BN have been shown to be stable and operate at elevated temperatures[55]. A red shift of the ZPL energy and ZPL broadening was seen with increase in temperature to 800 K, with both found to be reversible upon cooling back to room temperature[55]. The shift in ZPL energy upon heating and cooling was tentatively assigned to the expansion of substrate lattice upon heating and electron−lattice interactions[55]. The stability of SPE from h-BN at elevated temperature is suggestive of the possible uses in integrated quantum technologies for real-world environments. Another recent work, which investigated the temperature dependent line width, line shift, excited state lifetime, and intensity of quantum emitters in h-BN has also reported similar red shifts and ZPL broadenings for both Group 1 and Group 2 emitters upon increasing the temperature[111]. Piezoelectric coupling to the low energy acoustic phonons was proposed to explain the exponential line width broadening of the ZPL[111] with temperature.

**2.5 SPE intensities, quantum yields, radiative lifetimes, and spectral line widths**

A wide range of emitter intensities are observed in experiments[33-34, 38]. This could arise owing to weak absorption or else through loss of quantum yield either: purely by non-radiative means through other defect states, by non-radiative means to some lower-energy radiative state, or by energy transfer to some neighbouring defect. Related to this are the radiative lifetimes, which could be controlled by the intrinsic oscillator strength of the SPE or else by energy-loss mechanisms. Group 1 and Group 2 SPEs have different radiative and non-radiative lifetimes, indicating both key differences in internal electronic structure and in their interaction with neighbouring impurities, possibly owing to the proximity of a centre to the surface being different[33]. In other experiments[113], radiative lifetimes have been measured and correlated to coherence times apparent from spectral line widths. The results indicate that linewidths can be much broader than those anticipated from the radiative lifetimes, indicating some inhomogeneous broadening process. Further, this inhomogeneous broadening can change as a function of time and the number of times a defect emits a photon. This has been interpreted in terms of charge or other fluctuations in the h-BN environment influencing SPE properties through a Stark shift[113].

## 2.6 Interaction with phonons

As exampled in Fig. 1 and summarized in Table 1, the PL spectra from defects in h-BN reported in different studies have reorganization energies in the range of 15 – 30 meV for Group 2 and 60 – 160 meV for Group 1. These values are all quite small, with, for comparison, excited states of chemical defects considered in DFT and other computational studies[88] being predicted to have reorganization energies up to 3000 meV. Nevertheless, the reorganization energies of even these low magnitudes are significant, leading to observed Debye-Waller factors of 15 – 40 % for Group 1 and 50 – 80 % (or perhaps even more) for Group 2. At high temperatures, phonons also contribute to the spectrum through hot-band anti-Stokes transitions[61] and associated inhomogeneous broadening effects[113].

## 2.7 Absorption and emission dipole polarisation

In this section, we first discuss the effect of dipole allowed spin-preserved transitions. In lowest order, such transitions may happen via interaction of the electric field vector of the light with the electric transition dipole of the defect,

$$\mathbf{d}_{if} = \langle \Psi_i | \hat{\mathbf{d}} | \Psi_f \rangle,$$

where $|\Psi_i\rangle$ and $|\Psi_f\rangle$ are the electronic wavefunctions of the initial and final states, respectively, and $\hat{\mathbf{d}}$ is the electronic dipole operator. This is a vector quantity that will be aligned with respect to defect orientation and so will respond differently to incoming light of different angles and orientations to the h-BN plane and to light of different polarisations. Emission of light will also occur in specific directions with specific polarizations. If defects have local point-group symmetry, then only certain orientations and polarizations of the absorbed and emitted photons will be allowed. Within the Huang-Rhys model, direct absorbers (Fig. 4(a)) must have their absorption and emission directions and polarizations aligned. Beyond the Huang-Rhys model, direct absorbers may absorb and/or emit light in additional polarizations owing to the Herzberg-Teller effect[97], which vibronically couples the electronic transition to phonon transitions in vibrational modes than can be non-totally symmetric. This effect leads to "false origins" that appear akin to ZPL's of other defects except that they occur at energies different from that of the ZPL according to the vibration frequency of the coupled mode. Beyond the direct-absorption model, indirect absorption can occur to a state with a different polarization to the state that eventually produces SPE, leading to a misalignment of absorption and emission polarisation.

The emission and absorption dipoles for h-BN emitters have been found to be aligned when excited within 160 meV of the ZPL, but can be misaligned otherwise[34]. A threshold of 160 meV is very close to that expected based on the Herzberg-Teller effect, but as SPE persists for excitation at much higher energies than this, it has been concluded that both direct and indirect excitations are responsible for the polarization changes[34]. The existence of a fast decaying intermediate and a long-lived metastable state accessible from the first excited electronic state has also been proposed in another recent study[63]. Indirect absorption mechanisms supporting misalignment of emission and absorption dipoles has been envisaged in another recent study[62]. Polarized in-plane emission is observed after in-plane polarized light absorption[34]. In h-BN crystals, this emission is not necessarily correlated with crystal lattice directions, implying that the defects do not have high symmetry[34]. Also, the absorption can be either aligned parallel to the emission or else at some angle to it. Both types are observed, with no particular misalignment angle preferred[34]. In principle, this could arise following, for one polarization, Franck-Condon allowed[96] transitions, and for the other polarization, Herzberg-Teller-allowed[97] vibronic transitions, all within the same electronic state[114]. However, the wide and variable energy spacing between the absorptions of different polarization imply that instead two states are involved, supporting Franck-Condon allowed, differently polarized, in-plane absorptions by indirect paths such as those shown in Fig. 4(b)-(c).

## 2.8 Other evidence suggesting that more than two states are often involved in SPE function.

Many other experiments have been interpreted as showing that the simple direct-excitation model (Fig. 4(a)) is not always adequate. Bright single-photon emission with high quantum efficiency is achieved only if the excitation wavelength is matched to the emitter, providing resonant excitation[63], suggesting that higher-energy excitation facilitates alternative processes and hence a complex level structure of the culprit defect. Another study reports the presence of possible intermediate states by utilizing different laser powers to create two-photon absorption[56]. Complex behaviour observed in antibunching experiments is also suggestive of a complex state structure of emitters in h-BN. This is also supported by another study, which reports large ZPL spectral diffusion and discrete jumps of up to 100 nm after exciting the emitters with blue lasers, while only a fraction of emitters fluorescing on green illumination[115]. This is interpreted in terms of the initially excited state being able to emit directly (Fig. 4(a)), in competition with internal conversion to a third state that also emits, but at a much lower energy. Such a scenario suggests the possibility of addressing different defect states or levels[115]. Since in quantum optics it is highly desirable to have the ability to address different defect levels, defects in h-BN appear attractive.

Subsequent studies[62] on a Group 2 emitter were interpreted in terms of Fig. 4(b). Initial absorption was taken to be followed by intersystem crossing and PL by phosphorescence. The most plausible explanation of the data was taken to be absorption within the singlet manifold followed by intersystem crossing to a triplet state, but the related possibility of absorption within the triplet manifold followed by intersystem crossing to a singlet state could not be excluded; transitions between doublet and quartet states were considered unlikely, however. Subsequently, related defects have been demonstrated to produce photoluminescence contrast using ODMR techniques[70, 75], unambiguously revealing triplet ground-states.

## 2.9 Strain tunability

Spectral tunability of h-BN single photon emitters has been achieved over an energy range of 6 meV[42]. Combined with the ability to address different defect levels, this mechanical control over the emission wavelength allows for considerable emission control. Indeed, such tuning greatly increases the probability of assembling different SPEs together with spectral overlap. Although in this study[42], the emission dipole was in the plane of h-BN sheet, in general spin-preserved transitions can have their dipole in either the plane or perpendicular to the plane of h-BN sheet.

Another recent study reports anomalous pressure dependent shift of the emission lines from h-BN emitters with three different behaviours *i.e.* a red shift (negative pressure coefficient), a blue shift (positive pressure coefficient), or even a sign change from negative to positive[54]. This behaviour is suggestive of existence of competition between different interlayer interactions leading to such anomalies.

## 2.10 Stark effects

The electrical control of single-photon emission from atomic defects in *h*-BN has been recently demonstrated via the Stark effect[116]. Stark shifts were measured to be of the order of 5.4 nm per GV/m, with comparable values anticipated by calculations of model defects[116]. Fits to the measured data gave small dipole moment changes on excitation of $\Delta \mathbf{d} = |\mathbf{d}_{ff} - \mathbf{d}_{ii}|$ for the defects of less than 1 D, with large analogous polarizability changes of about 150 A$^3$. Another recent experiment reports a relatively large and reversible Stark shift of 5.5 ± 0.3 nm at a zero-field wavelength of 670 nm induced by applying 20 V across the sample[117]. These measurements were conducted by applying an electric field perpendicular to the substrate and are therefore explained by out-of-plane distortion of the defect. In-plane electric fields could give rise to much larger energy shifts and such giant Stark effects of an SPE have recently been reported[118]. The orientation of the electric permanent dipole moment in that solid-state SPE was determined via angle-resolved Stark effect measurements, revealing intrinsic broken symmetry at the colour center[118].

## 2.11 Magnetic states for quantum device applications

For applications such as ODMR[74], the role of the emitter is primarily to facilitate initialization and readout of individual electron spin sub-states[119]. For ODMR applications, there is no requirement for single-photon emission per se, only that different recombination pathways provide spin dependent contrast in the emission intensity. In this section, we review magneto-optical studies on SPEs in h-BN. Traditionally, non-magnetic SPEs involving singlet ground states, possibly intrinsic to a single defect or else from spin-coupled adjacent paired defects, have been associated as possible Group 2 emitters[37, 112]. A recent study also reported no change in the optical spectra of h-BN defects with the application of a magnetic field[112]. However in another recent work, anisotropic PL patterns were observed as a function of applied magnetic field for some selected Group-2 emitters, with other similar emitters in the same sample showing no change[62]. Further, systematic changes in brightness with applied magnetic field for some emitters in h-BN is suggestive of optically addressable spin and hence a possible mechanism of spin-initialization and readout[62]. In this work[62], adsorption and emission was interpreted as arising within the singlet manifold, with intersystem crossing occurring also to the triplet manifold. However, recent ODMR studies[70, 75] probing defects indicate absorption and emission in the triplet manifold with intersystem crossing to a singlet state. The ODMR emitters may also characterised by unusually weak and unprecedently broad emission[70], or else the presence of many discrete Group-2-like emitters in a small energy range[75] and the involvement of carbon-atom impurities. Magnetic ground states have also been reported in samples following neutron irradiation[112].

**2.12 Hyperfine Coupling**

Another method used for defect identification is electron paramagnetic resonance (EPR) spectroscopy. Unfortunately, it is difficult to relate observed EPR signals to observed PL, and, in principle, different types of defects could be responsible for these different signals. This problem is overcome in the ODMR experiments described before[70, 75], but otherwise EPR responses have been obtained[120-121] following the preparation of single layer h-BN, subsequent to the deterministic production of defects through electron beam radiation[43], and following carbon doping; of particular interest is that carbon-doped centres can give rise to intense PL in the UV region[52, 122].

Independent of their source, paramagnetic centres can have considerable effects on the electronic and magnetic properties of h-BN. Based on interpretation of the EPR data, both nitrogen ($N_v$) and boron ($B_v$) vacancies have been identified as paramagnetic centres[123], with $N_v$ being more prevelant[124-129]. Two types of $N_v$ paramagnetic centres have been identified: (i) three-boron centres (TBC), in which an unpaired electron interacts with three equivalent boron ($B^{11}$) nuclei, producing 10-line EPR spectra, and (ii), one-boron centres (OBC), in which oxidative damage at the centre forces the unpaired electron to interact with only a single $B^{11}$, producing 4-line EPR spectra. The TBC can be deliberately produced either by irradiation[124-

[126] or by carbon doping[128], but controllable h-BN oxidation to produce OBCs has not yet been achieved.

ODMR measurements could provide collaborating evidence to support these EPR assignments. In one ODMR study[70], the defect was tentatively assigned to a negatively charged boron vacancy ($B_v^{-1}$) possessing a high-spin triplet ground state[70]. In this study, EPR and PL techniques were employed to determine the parameters of a spin Hamiltonian depicting the defects involved[70], with the hyperfine coupling parameter for this defect thus determined to be 47 MHz[70]. In another ODMR study[75], reported a related quantum-emitting source with a hyperfine coupling parameter of around 10 MHz, with its angular dependence indicating an unpaired electron in an out of plane π-orbital[75]. The ZPL energies for these two ODMR studies are distinct, indicating different defects are responsible.

## 2.13 Zero field splitting

As mentioned in the previous section, some recent experimental studies have proposed, based on EPR, ODMR and PL measurements, defects centres in h-BN with high-spin triplet ground states ($S = 1$)[70, 75]. In general, a $S = 1$ electron spin system can be described by a spin Hamiltonian of the form

$$H = g_e \beta \boldsymbol{B} \hat{S} + \hat{S} D \hat{S}.$$

Here $g_e$ is the electronic g-factor ($g_e = 2.0028 \pm 0.0003$), $\boldsymbol{B}$ is the external magnetic field and $D$ is the zero-field splitting tensor. This tensor comprises the anisotropic dipolar interaction of the two electron spins forming the triplet state, averaged over their wave function. This tensor is traceless and thus characterized by two parameters, $D$ and $E$, known as the axial and rhombic zero-field splitting parameters, respectively. This spin-spin interaction arises owing to the non-spherical shape of molecular orbitals lifting the degeneracy of multiplets. In a recent study, a defect species with a triplet ground state has been reported with $D = 1.2$ GHz[130]. The two recent ODMR studies report emission in h-BN in one case with a ZFS splitting parameter of 3.6 GHz[70] and in the other no detectable ZFS indicating an upper bound on the ZFS splitting parameter of 4 MHz[75].

## 2.14 Overview of the effect of h-BN hosts on the formation of SPEs

SPEs have been realized in different h-BN host materials including monolayers[1], thin films[38], bulk crystals of h-BN[35, 131], tap-exfoliated[132,54] and liquid exfoliated nano-flakes[40], nanotubes and other nanostructures[112,110, 133], h-BN samples often require post growth processing, e.g., annealing or irradiation by ions and low energy (keV) or high energy (MeV) electrons[134,135], femtosecond-pulsed laser irradiation[132] or plasma processing[136], etc., for stable emitters to be observed, and it is not clear whether the post growth treatments create emitters or just activate pre-existing defects. Defect activation seems more likely than creation because all attempts at engineering emitter arrays suffer from low success rate probabilities[73].

Nevertheless, certain patterns in the formation of SPEs in different hosts are noteworthy. Some studies report preferential occurrence of emitters at or near grain boundaries and flake edges[132,54], which is undesirable for device applications. Indeed, Group 2 emitters have been identified near structural defects in the basal plane of the h-BN[37]. However, this is not a universal characteristic, and emitters do also occur well away from edges and grain boundaries[134,38]. It is a general observation that the CVD-grown large-area h-BN films can host a high density of SPEs that are distributed uniformly throughout the entire area of the film[137]. However, the purity of these emitters has been low, until lately, hindering their applications in practical devices[73]. Very recently, efficient post-growth processing of h-BN films transferred from substrates used for chemical vapour deposition (CVD)[138] has been developed, significantly improving SPEs in h-BN.

Recently, X-ray photoelectron spectroscopy (XPS) and PL studies have been performed on h-BN grown by mechanically exfoliating the bulk crystal, and also CVD-grown layers on copper foil, with the extracted layers then transferred onto $SiO_2$ substrates[139]. The exfoliated h-BN did not show the presence of bright emitters, whereas the CVD-grown sample showed a much brighter background and photo-stable emission at specific locations. The optoelectronic properties of h-BN between the samples are thus inherently different, and so are their binding profiles. The boron binding energy peaks for the CVD-grown h-BN were found to be broadened and shifted with respect to that corresponding to the exfoliated h-BN[139]. These changes suggest that part of the new boron-related bonding present in the CVD sample is responsible for the observed differences in emission and optoelectronic properties[139].

We conclude this section on the following note: in principle, a comparison of properties of SPEs in different hosts can be useful in understanding the formation and structure of the defect. However, the effect of post-growth treatment on different h-BN samples need to be understood in detail before drawing any conclusions.

**2.15 Summary**

Based on early studies, SPEs in h-BN have been categorized into two primary classes, Group 1 and Group 2, but we see now that this classification is inadequate. For example, most SPEs have been classified as having non-magnetic ground states[37, 112], but advances this year have brought important observations considering the detection and utilization of Group 2 defects with magnetic ground states[62, 70, 75, 130]. The search for the understanding of the chemical nature of h-BN defects is therefore not just the search for a single iconic defect, but rather the search for multiple forms with some similar and some contrasting properties.

**3. Theoretical and computational modelling**

In this section, we briefly discuss the theoretical attempts made so far to reproduce the observed properties of SPEs in model defect systems. These studies have majorly focussed on possible point defects in h-BN[85-88, 90, 140-142]. Initial studies were crude in nature, allowing an overview to be determined concerning many postulated defects[85]. This work then spurned many subsequent studies considering each only one or else a small number of candidates[54, 84, 86-88, 116, 140, 143], focusing on the manifold of states available to each defect and on making reliable predictions. Of the observed defects, focus was placed on not only the chemical structures of common Group 1 and Group 2 emitters associated with non-magnetic ground states, but also on the rare but technologically important paramagnetic emitters seen in EPR experiments that are relevant to ODMR.

In the initial report[1] of SPE in h-BN, crude DFT calculations were undertaken in an attempt to identify the defect responsible. These calculations were performed within the generalized gradient approximation using the PBE density functional, taking no account for the reorganisation energy in the final state. The nitrogen anti-site defect was identified as the most likely candidate. A subsequent and more thorough study by Abdi et al[87], using the HSE06 hybrid functional combined with the delta-SCF method (in which a single excited-state orbital occupation is enforced during the calculation), showed that the error introduced by the PBE functional is almost exactly cancelled by the neglect of the reorganisation energy.

In the original survey study[85] overviewing a large number of likely defects, calculations were based again on the PBE density functional[144] and the delta-SCF method. Only electronic states of neutral defects involving two deep defect levels (Fig. 3) were considered. Magnetic properties were also calculated. The result was that a set of 13 possible defects were shortlisted[85]. For each of these the Huang-Rhys factors were evaluated and the PL spectral intensity distributions simulated. The net result was the identification of three defects of particular interest: $V_NN_B$ (a nitrogen vacancy and one of the surrounding boron replaced with a nitrogen), $V_NC_B$ (nitrogen vacancy and one of surrounding boron replaced with a carbon) and $V_BC_N$ (boron vacancy and one of surrounding nitrogen replaced with a carbon).

We consider the predictions initially made[85] concerning $V_NC_B$. This defect had a triplet ground state, as is relevant to observed Group-2 paramagnetic defects, and the smallest calculated Huang-Rhys factor for any considered defect of $S^E = 1.66$, with associated $\lambda^E = 180$ meV, for the $(2)^3B_1 \rightarrow (1)^3B_1$ emission of $V_NC_B$. However, such large Huang-Rhys factors and reorganization energies would appear to be relevant only to Group-1 emitters. Hence all defects considered therein naively appear to be only relevant for Group 1 emitters, but given the very small magnitude of all observed reorganization energies in absolute terms, calculations made at the used level of theory would not be expected to be sufficiently reliable to make such a conclusion. A more reliable measure is in fact the comparison of observed and calculated spectra, and the agreement was also found to be poor, as reproduced in Fig. 5[85].

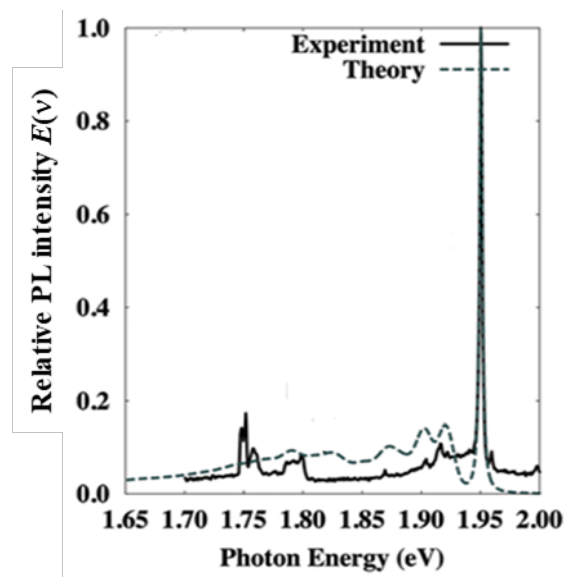

**Fig. 5** Theoretical prediction for $V_NC_B$ of its $(2)^3B_1 \rightarrow (1)^3B_1$ emission, compared to an observed Group-1 luminescence line shape from an h-BN defect, after alignment of the observed and calculated ZPL energies. Adapted with permission from[85].

Extensive high-level calculations were then performed for the $V_NC_B$ defect[86, 88], considering 25 possible states of the defect selected on the basis of their adiabatic transition energy $\Delta E_0$. These predicted a singlet ground state instead of triplet, but did preserve a low-bandwidth triplet emission following any intersystem crossing (Fig. 4(c)). Also, subsequent calculations[116] predicted that the singlet ground state of $V_NC_B$ distorts to a non-planar asymmetric structure with a reorganization energy of 440 meV, predicting large energy dissipation on the ground-state recovery reaction (Fig. 4(c)), inconsistent with observed energetics. Any role for $V_NC_B$ therefore now appears unlikely.

Recently, boron-dangling bonds have been proposed, based upon the energetic positions of the states within the band gap [90], as the likely source of the observed single-photon emission around 2 eV. Such dangling bonds are postulated to occur inside voids in the h-BN that have become partially hydrogen terminated (envisaged from h-BN, e.g., by removal of 3 boron and one nitrogen atoms and the addition of four hydrogens to leave one unterminated boron dangling bond). In this defect, the emission was predicted to be linearly polarized with the absorptive and emissive dipole aligned[90]. However, the existence of defect orbitals close to the bulk bands (conduction band) may make this assignment inconsistent with experimental observation of stable single photon emission even at elevated temperatures[55].

Another two recent theoretical studies propose $V_B^{-1}$ [141-142] and $V_N^{-1}$ [141] as the source of observed ODMR signals[70] from h-BN. The detailed magneto-optical properties and corresponding radiative and non-radiative routes which are responsible for the optical spin polarization and spin dependent luminescence of the defects are modelled in these studies[141-142]. Further detailed studies focused on calculating the dipole strength of different spin preserved transitions, ISC rates, coupling to phonons, benchmarking of DFT energies etc., are however, needed to authoritatively assert the precise nature of the emitters generating some ODMR signals[70]. A critical role for carbon-atom defects has been reasoned for other observed ODMR signals[75].

Even without knowing the precise chemical nature of the different types of h-BN defects, DFT can still be useful in depicting generic expected SPE properties. For example, general aspects of spectral tuneability through the applied strain have been backed up by DFT simulations[42] on the proposed $V_NN_B$ defect, as shown in Fig. 6. These simulations considered four strain directions along the plane of the h-BN sample, as labelled by the lattice directions shown in the inset in the figure. A realistic case, similar to experimental conditions, is taken into account by considering the effects of the Poisson's ratio of 0.37 of the bendable polycarbonate substrate. Under these conditions, tensile strain along the AC2-direction produces a compression along the orthogonal direction ZZ1: $\varepsilon_{ZZ1} = -0.37 \times \varepsilon_{AC2}$, and vice versa. The simulations confirm the observed non-monotonic behaviour of the ZPL energy under the effect of strain and its role in the large spectral distribution[42].

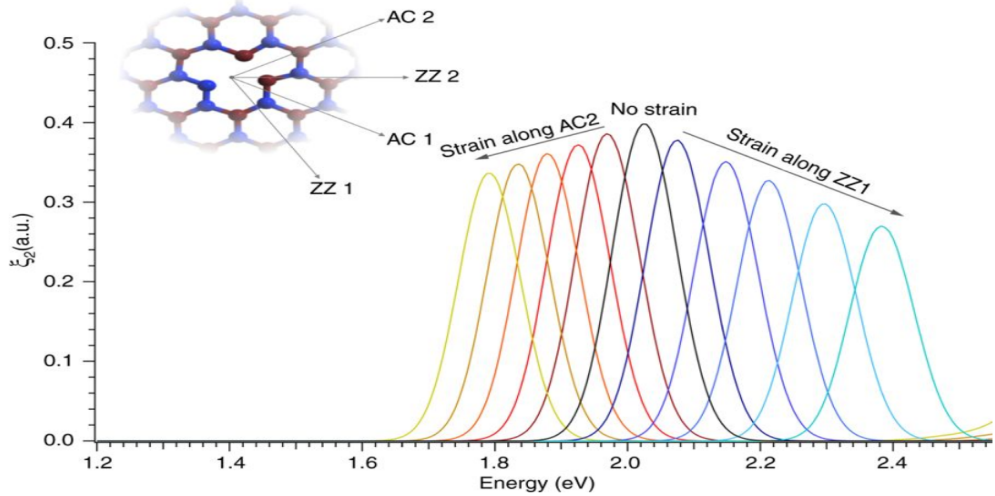

**Fig. 6** Simulated absorption spectra as a function of strain[42]. The cartoon in the inset shows the lattice directions in the plane of hexagonal lattice with respect to the $V_NN_B$ defect geometry. Four independent directions can be identified: two along zigzag directions, ZZ1 and ZZ2 and two along the armchair directions, AC1 and AC2. The main plot is simulated optical response in the form of imaginary part of dielectric function ($\xi_2$), a quantity proportional to $A(\nu)$, as a function of the strain applied in either the zigzag or armchair directions. The black graph shows the absorption at zero strain. The graphs in red (blue) tones correspond to tensile strain applied along the AC2 (ZZ1) direction from 1 to 5%. Reprinted with permission from ref.[42]

A-priori calculations using DFT can provide useful tools for the interpretation of EPR data such as observed zero-field splitting parameters $D$ and/or $E$ [21-22, 86], as well as also hyperfine tensors[21-22]. Indeed, such calculations have been helpful in the identification of point defects in different semiconductors by comparing observed and calculated hyperfine constants[28-22, 145-148]. Moreover, it has been shown that DFT can calculate spin-related quantities of semi-conductor point defects e.g. hyperfine tensors and zero field splitting with considerable accuracy[149].

In our recent work, we have assigned observed h-BN EPR signals at 22.4 MHz (TBC type), 20.83 MHz (TBC type), and 352.70 MHz (OBC type) to the $V_N$, $C_N$, and $V_NO_2$ defects, respectively[86]. However, there is no evidence that these EPR-active defects also account for any observed single photon emission. In addition, we have calculated the HF values for various carbon-related anti-site defects, in which a nitrogen or boron vacancy is accompanied by a neighbouring carbon-atom and/or silicon-atom substitution. It is of interest to see if these, hopefully robust, predictions can be attributed to observed EPR features. Indeed, a recent EPR study has been performed taking on this type of challenge[130]. Broad optical absorption bands and PL bands centred at ~490 nm and 820 nm, respectively, have been assigned tentatively to defect species with a triplet ground state[130]. Another two experimental studies[75-70] have reported defect species in h-BN with triplet ground states. These studies should be backed up with further DFT calculations to assign the observed signal to particular h-BN defects.

Also, DFT calculations can be used to assign observed core-level binding energies, with, e.g., the boron $B_1$ peaks observed from CVD-grown h-BN samples attributed to either the $V_N$ or $V_NN_B$ defects[139]. Unfortunately, owing to resolution limitations and the weak signal amplitude the peaks could not therein be further resolved into the $V_N$ and $V_NN_B$ contributions[139].

## 4. Major theoretical/computational challenges

The precise chemical nature of observed defects is difficult to determine and hence first-principles computational techniques have been widely applied to aid in data interpretation[21-22, 85, 145-146, 148, 150]. However, methods for performing accurate calculations concerning PL for the types required are still being developed. The spectroscopy of relevant small molecules may be described to chemical accuracy using standard *ab initio* computational methods such as quantum Monte Carlo (QMC)[151], coupled-cluster singles and doubles (CCSD)[152] theory with perturbation corrections for triples excitations (CCSD(T))[153], related time-dependent approaches such as equations of motion coupled cluster (EOMCCSD)[154-155], singles and doubles multi-reference configuration interaction (MRCI)[156-157], as well as its approximation, complete-active-space self-consistent field (CASSCF)[158] theory with perturbative corrections for singles and doubles excitations (CASPT2)[159].

Whilst CCSD(T) is regarded as being the "gold standard" in molecular electronic structure computation, it is only quantitatively accurate when the ground-state can be well described using single-reference methods. This occurs when just one assignment of electrons into defect and/or host levels (Fig. 3) is adequate to describe electronic states (Fig. 4). This is rarely the situation when broken chemical bonds are involved, and isolated h-BN vacancies generate 6 defect levels pertaining to broken bonds. Hence defect states are usually multi-reference in character and embody strong static electron correlation, demanding QMC or MRCI approaches instead of CCSD-based ones. Nevertheless, dynamic electron correlation is likely also to be important, and its improved description in methods like CCSD(T) compared to MRCI presents its own advantage. Approaches like CCSD(T) do embody static electron correlation and converge to the exact answer with increasing treatment order, it is just that it is included asymmetrically, slowing convergence.

Mostly CCSD and MRCI based methods are only applied to model compounds[160], not materials such as 3-D h-BN or 2-D h-BN Nano flakes. This raises the question as to how accurately the model compounds reflect the real situation. A range of improved high-level methods are available[161-179], but unfortunately rarely deployed. Simply increasing cluster size until apparent convergence is reached may not be adequate as long-range dielectric properties could control localized defect energetics. That reasonable results for small cluster sizes could be expected is anticipated based on the very different natures of 2D and 3D dielectrics, with effects in 2D materials taking a different form that is nowhere near as significant as are the well-known effects of dielectric screening in 3D materials[69].

Nevertheless, the most commonly applied approaches to electronic and nuclear structure simulation in modern times is DFT and its time-dependent variant, time-dependent DFT (TDDFT). Despite having no method available for the systematic improvement of DFT calculations towards the exact answer, in practical calculations on sizeable molecules, DFT and TDDFT approaches can often deliver similar accuracy to their much more computationally expensive *ab initio* counterparts. Most DFT approaches in common use treat multi-reference character and static electron correlation only poorly, however, making their application to defect spectroscopy unusually difficult; a range of more pertinent new approaches are being developed, however[180]. On the other hand, DFT is regularly applied to periodic systems and does not require cluster models. In molecular-cluster-based approaches, the size of the cluster must be increased until all long-range effects of interest are included, a difficult task if transitions involve the valence and/or conduction bands of the solid or dielectric screening is important, whereas for periodic-solid approaches, the size of the unit cell must be increased until neighboring defect sites do not interfere with each other. Either way presents computational challenges.

For the interpretation of results, an important feature is that DFT and TDDFT can suffer various severe failings depending on the type of density functional used and the nature of the state being considered,[181-187] demanding that great care be taken. Efficient and widely used DFT functionals like PBE[144] underestimate band gaps and so place what should be deep defect levels (Fig. 3) near or inside the VB or CB, causing defect states to delocalize over the sample rather than localize on the defect. The effect also results in poor predictions of ZPL energies, which ideally should have 0.1 eV accuracy, although 0.4 eV would be a useful level of accuracy. This is only possible if both static and dynamic electron correlation effects are taken into consideration.

As speed and efficiency was critical, the initial DFT calculations performed for survey various defect systems and calculation of PL lineshapes in ref.[85] were performed using PBE, a method known to underestimate semiconductor band gaps. As discussed earlier, in that study[85] $V_NC_B$ was proposed as the possible emitting centre on the basis of the calculation of the Huang-Rhys factor and reorganization energy for a plausible transition. However, owing to theoretical/computational limitations described above, this defect shows considerable complexity, with many unresolved issues concerning interpretation of experimental data[62, 85, 89, 150], and there is a considerable controversy in the literature regarding its electronic and geometric structure[116, 150].

Therefore, to calibrate and benchmark the DFT methods against *ab initio* quantum chemistry approaches, as well as to predict an accurate and complete optical cycle of $V_NC_B$ and another carbon related defect $V_BC_N$, later[88] we replaced PBE with the non-local Heyd-Scuseria-Ernzerhof hybrid functional (HSE06)[188-189] as an approximation to exchange and correlation[86]. We also performed[88] *ab initio* calculations using the CCSD(T)[152-153], EOMCCSD[154], and CASPT2[159] and MRCI[156-157] methods on a model compound for 25 states of $V_NC_B$, utilized the results to make realistic predictions of PL energies, as shown in Fig. 7(a).[88] Hybrid functionals like HSE06 were found to work very well for excitations within the triplet manifold of the defect, with an accuracy equivalent to or perhaps exceeding the accuracy of the *ab initio* methods used. However, HSE06 underestimates triplet state energies by on average 0.7 eV compared to closed-shell singlet states, while open-shell singlet states are predicted to be too low in energy by 1.0 eV, as shown in Fig. 7(a). This leads to miss-assignment of the ground state of the $V_NC_B$ which was predicted to be triplet in character[150], whilst we found that the singlet ground state is lower in energy compared to the lowest triplet state by ~1 eV, after applying the correction factors. Therefore, similar approach should be adopted on other defects to have an energetically useful comparison with experiment. It is not certain how transferrable such corrections will prove to be, owing to the large variability in energy-level scenario and occupation that different defects and charge states present. Indeed, a similar benchmarking approach has also been adopted for correcting the energies of electronic states of $V_NN_B$[84] and corrections quite different to $V_NC_B$, as shown in Fig. 7(b).

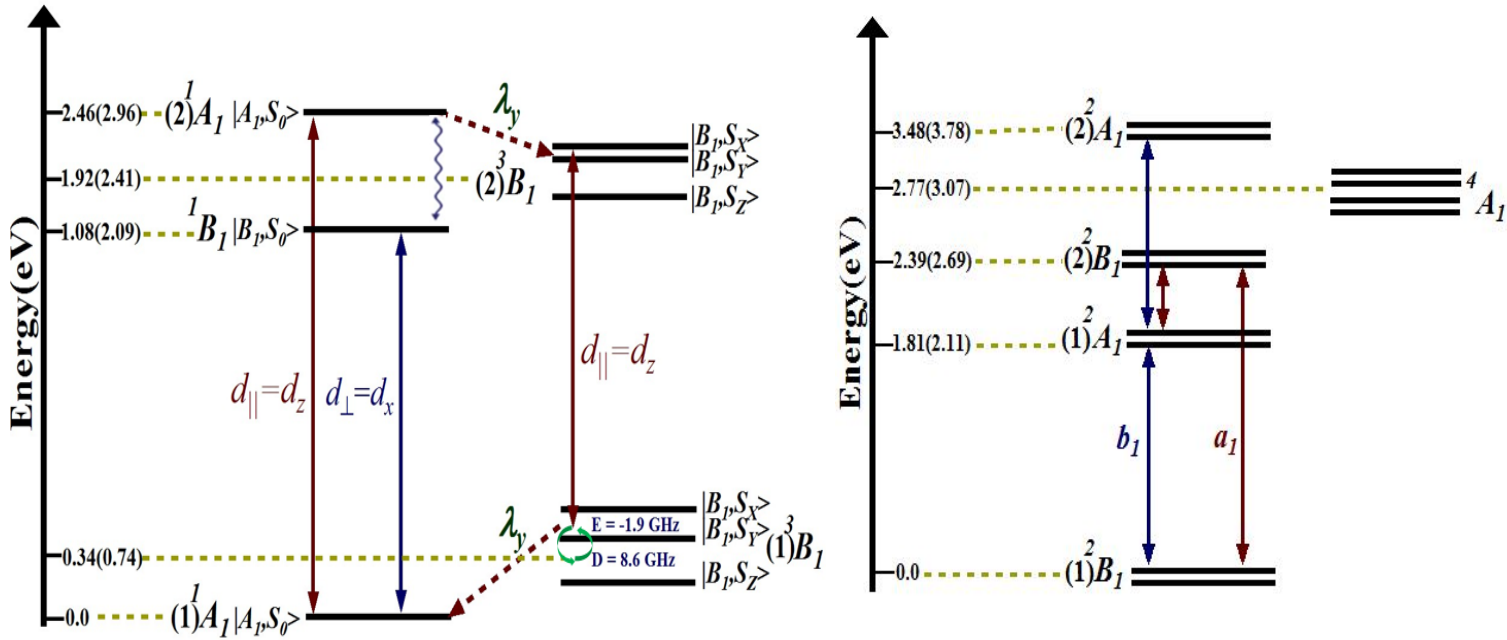

**Fig. 7** HSE06 adiabatic energies of low lying states of (a) $V_NC_B$[86, 88] and (b) $V_NN_B$ as calculated by DFT, with, in (), these energies corrected according to *ab initio* CCSD(T), EOMCCSD, and CASPT2 calculations for a model compound. Allowed transition polarizations *d*, spin-orbit couplings λ driving non-radiative transitions and zero-field splittings are also indicated. Reprinted with permission.[86, 88]

Concerning the choice of density functional used in DFT calculations, comparison of the calculations with simple generalized-gradient functionals like PBE revealed their failure at the most basic level and should not be applied when quantitative analysis of defect states is required[88]. While HSE06 appears to be much more useful, we also found[88] that the long-range corrected hybrid functional CAM-B3LYP[186, 190-191] performed much better than HSE06 for calculations on model compounds compared to *ab initio* approaches, but unfortunately it has limited availability in software packages supporting periodic boundary conditions.

It is worth mentioning here that spin-restricted Kohn-Sham (RKS) scheme in which the spin-up and spin-down electrons are described using the same orbitals fails at large bond distances, converging to species with strong ionic contamination at high energies that lead to the incorrect determination of the magnetic properties of the asymptotic ground and excited states[192]. Therefore, the spins should be allowed to adopt different orientations in variational space, i.e., the spin-unrestricted Kohn−Sham (UKS) scheme, which facilitates disassociation to the correct ground state[192]. This procedure works when two electrons are distributed amongst two orbitals as in a traditional single covalent bond, but more complex scenarios may easily arise for defect states for which it will be found inadequate.

It is clear from this discussion that the correction factors and hence the computational challenges when it comes to dealing with h-BN defects are system, defect, and state specific. In other words, the electron correlation effects are quite different for each defect system and for each state depending upon the localization of the states and therefore a generalization cannot be made.

Another approach for estimating the likelihood of different defects occurring is through consideration of defect formation energies[89, 140]. This focuses on the probability of formation of a particular defect species under thermal equilibrium. Computationally, it is technically difficult to apply as standard methods for the calculation of defect formation energies in 3D materials cannot be applied to defects in 2D materials like h-BN. Owing to the divergence of the Coulomb energy with vacuum size, the results for the formation energy of defects can vary widely with supercell size[193]. Also, defect formation would most likely be under kinetic control rather than thermodynamic control as defects are intrinsically non-equilibrium in nature. Atoms are likely to come together during the formation of material and bind into a high energy state, with the barriers to further reaction being so large enough that the defect stays trapped there. Indeed experimentally identified defect species in SiC, InN etc. have high formation energies[194].

## 5. Conclusions

Point defects in h-BN are attractive owing to their highly efficient, stable, bright and linearly polarized room temperature SPE. The identification of the geometric, electronic and magnetic structures of the defect can be a next step toward their potential utilization in quantum technologies. Defects have been observed with wide-ranging properties, and it is clear that many types of defect contribute to the general phenomenon. Indeed, different defect compositions may find applications in different areas. In particular, for ODMR applications, a singlet-triplet system involving intersystem crossing could support the observed optical cycle and other experimental observations made for these SPEs.

It is possible that high-level calculations can contribute to the chemical identification of defects, but this is hampered by the diverse range of possibilities and the great challenges posed by defect spectroscopy, features that historically have not been the focus of algorithm and software development. To aid theory, one of the most important aspects of experimental research is the accurate measurement of the reorganization energies associated with SPE.

Regularly determining and reporting the PL bandshape beyond the 0-1 part of the PSB is important to this, dealing with experimental limitations arising from the $v^3$ (or $v^5$) dependence of the intensity in these critical regions. Reorganization energies are relatively easy to deduce from calculations as they do not require the determination of vibrational modes and frequencies or extensive spectral simulations. Having extensive experimental data available would therefore allow many more states and defects to be examined computationally using available resources. Also, the accurate measurement of spectral polarization is important, especially if it can be related to crystallographic axes of the h-BN. Related is the ability to simultaneously measure both PL and magnetic properties for a single defect. Characterisation of spectral polarization with respect to the axes of the host 2D or 3D material is also important, along with accurate Stark measurements. Modern approaches to Stark spectroscopy employing extremely large AC electric fields could prove useful[195]. An important property not yet measured is the absorption spectrum of defects, possibly accessible via fluorescence excitation measurements. These are difficult experiments owing to the low absolute intensity of SPEs, but would be very revealing, facilitating for example comparison of high-resolution absorption and emission data and hence determination of the Duschinsky matrix, a feature that could prove to be a controlling element in understanding PL, as it is known to be in natural photosystems[196-197].

The most important aspect of any calculation is to understand which predictions are likely to be reliable and which are not. Only conclusions for which the reliability can be estimated should be presented. Choice of sample used to model each considered defect, choice of computational method, and choice of excited states to consider are always critical. Certainly, any electronic-structure method used needs to be thoroughly benchmarked. However, for predicting spectral properties, it is not just the electronic structure method that is important, but also choice of approximation, including: the Born-Oppenheimer approximation, the Condon approximation, the assumption of harmonic potential-energy surfaces, the treatment of the Duschinsky matrix and vibrational frequency changes induced by PL, the temperature of experiments to be modelled, the influence of relaxation processes, energy transfer between defects, spectral diffusion, etc.

**ACKNOWLEDGMENTS**

SA acknowledges receipt of an Australian Postgraduate Award funded by ARC DP 150103317. JRR acknowledges support from the National Natural Science Foundation of China through Grant No. 11674212.

**AUTHOR INFORMATION**

**Corresponding Author**
* Sajid.Ali@alumni.uts.edu.au